\documentclass[sigconf]{acmart}

\usepackage[most]{tcolorbox}
\usepackage{xcolor}
\usepackage{fontawesome5}

\definecolor{declcolor}{RGB}{32,99,155}
\definecolor{intentcolor}{RGB}{46,125,50}
\definecolor{goldcolor}{RGB}{191,144,0}
\definecolor{boxgray}{RGB}{248,249,250}

\newtcolorbox{promptbox}[2][]{
    enhanced,
    breakable,
    colback=boxgray,
    colframe=#2,
    boxrule=0.8pt,
    arc=3mm,
    left=6mm,
    right=6mm,
    top=4mm,
    bottom=4mm,
    fonttitle=\bfseries\small,
    title=#1,
    attach boxed title to top left={
        yshift=-2mm,
        xshift=4mm
    },
    boxed title style={
        colback=#2,
        colframe=#2,
        arc=2mm,
        boxrule=0pt
    }
}

\newtcolorbox{promptboxf}[2][]{
    enhanced,
    breakable=false,
    colback=boxgray,
    colframe=#2,
    boxrule=0.8pt,
    arc=2mm,
    left=3mm,
    right=3mm,
    top=2.5mm,
    bottom=2.5mm,
    fonttitle=\bfseries\footnotesize,
    title=#1,
    attach boxed title to top left={yshift=-2mm, xshift=3mm},
    boxed title style={colback=#2, colframe=#2, arc=1.5mm, boxrule=0pt}
}

\newcommand{\TLA}{TLA\textsuperscript{+}\xspace}
\newcommand{\Ours}{\textsc{TLA\textsuperscript{+}-Bench}\xspace}

\definecolor{loyolamaroon}{RGB}{95,0,20}

\definecolor{subheadgrey}{RGB}{238,239,241}

\definecolor{yesgreen}{RGB}{22,135,71}
\definecolor{nored}{RGB}{197,48,48}
\definecolor{partialamber}{RGB}{201,138,4}
\newcommand{\yes}{\textcolor{yesgreen}{\faCheckCircle}}
\newcommand{\no}{\textcolor{nored}{\faTimesCircle}}
\newcommand{\partialx}{\textcolor{partialamber}{\faAdjust}}
\usepackage{booktabs}
\usepackage{multirow}
\usepackage{tikz}
\usetikzlibrary{positioning,fit,backgrounds}
\usepackage{xspace}
\usepackage{colortbl}
\usepackage{array}
\usepackage{makecell}
\usepackage{graphicx}
\usepackage{listings}
\usepackage{enumitem}
\usepackage{needspace}
\makeatletter
\g@addto@macro\UrlBreaks{\do\a\do\b\do\c\do\d\do\e\do\f\do\g\do\h\do\i\do\j\do\k\do\l\do\m\do\n\do\o\do\p\do\q\do\r\do\s\do\t\do\u\do\v\do\w\do\x\do\y\do\z\do\A\do\B\do\C\do\D\do\E\do\F\do\G\do\H\do\I\do\J\do\K\do\L\do\M\do\N\do\O\do\P\do\Q\do\R\do\S\do\T\do\U\do\V\do\W\do\X\do\Y\do\Z\do\0\do\1\do\2\do\3\do\4\do\5\do\6\do\7\do\8\do\9\do\.\do\-\do\_}
\makeatother
\lstdefinestyle{tla}{
  basicstyle=\ttfamily\scriptsize,
  numbers=left, numberstyle=\tiny\color{gray}, numbersep=6pt,
  breaklines=true, columns=fullflexible, keepspaces=true,
  frame=none, xleftmargin=14pt, aboveskip=2pt, belowskip=2pt,
  moredelim=[is][\bfseries\color{nored}]{@@}{@@}
}

\AtBeginDocument{%
  }

\copyrightyear{2027}
\acmYear{2027}
\acmDOI{XXXXXXX.XXXXXXX}
\acmConference[KDD '27]{The 33rd ACM SIGKDD Conference on Knowledge Discovery and Data Mining}{August 2027}{City, Country}
\acmISBN{978-1-4503-XXXX-X/2027/08}

\setcopyright{none}
\renewcommand\footnotetextcopyrightpermission[1]{}

\begin{document}

\title{\Ours: An Execution-Grounded Benchmark and Dataset for Natural-Language to \TLA Specification Generation}

\newcommand{\lucaff}{\affiliation{%
  \institution{Loyola University Chicago}
  \city{Chicago}\state{Illinois}\country{USA}}}

\author{Arslan Bisharat}\lucaff
\author{Eric Spencer}\lucaff
\author{Brian Ortiz}\lucaff
\author{Khushboo Bhadauria}\lucaff
\author{Mujtaba Nazari}\lucaff
\author{Beatriz Santos}\lucaff
\author{Anisa Ramos}\lucaff
\author{TaiNing Wang}\lucaff
\author{George K. Thiruvathukal}\lucaff
\author{Konstantin L\"aufer}\lucaff
\author{Mohammed Abuhamad}\lucaff

\renewcommand{\shortauthors}{Bisharat et al.}

\definecolor{tofillbg}{rgb}{1.0,0.9,0.2}
\newcommand{\tofill}[1]{\textbf{\colorbox{tofillbg}{XX}}}
\newcommand{\predict}[1]{\colorbox{tofillbg}{#1\,\textsuperscript{p}}}

\begin{abstract}
Large language models increasingly write TLA$^{+}$ formal specifications from
natural-language descriptions, but progress is hard to measure: existing resources grade by
resemblance to a reference or by whether the output parses, neither of which shows correctness. We
present \Ours, a dataset and benchmark that grades by execution. Every gold specification ships a
configuration the TLA$^{+}$ model checker runs over the full reachable state space,
deciding exactly whether the specification holds the properties that configuration names. The
dataset holds 403 model-checked gold and 897 parse-only silver specifications from 13 public
repositories, subsumes prior TLA$^{+}$ generation data, and carries four model-written
descriptions in two styles from two providers, with difficulty and category labels. Our main
finding is about measurement itself: an exact oracle gives not one correctness number but a range.
Varying only the grading choices earlier benchmarks leave unstated, on one fixed set of model
outputs, the correct rate moves sixfold, from 10.0\% to 1.7\%; adding the interface-supply choice,
where the model is told the configuration's names, widens the range to elevenfold, from 18.7\% to
1.7\%. We call this range the correctness envelope and measure each of its bounds. The findings inside it are stable. Every model
writes valid TLA$^{+}$ far more often than correct TLA$^{+}$: the
strongest is correct 16\% of the time by default and 26\% when given the interface names, open
models at most 1\%, and correctness falls sharply with difficulty.
\end{abstract}




\maketitle

\section{Introduction}
\label{sec:intro}

Formal specifications state what a system must do precisely enough that a tool can check them.
TLA\textsuperscript{+} is a widely used specification language for concurrent and distributed
systems. It comes with a parser and a model checker that decide, over the full reachable state
space, whether a specification holds its stated properties~\cite{lamport2002specifying}. A
growing body of work asks large language models to write TLA\textsuperscript{+} from a
natural-language description. That raises a basic question for the field. How do we measure
progress on this task in a way that reflects correctness rather than surface form? The
data-science community meets this question whenever a model turns a description into an
executable artifact, as in natural-language to SQL~\cite{nl2sqlbugs} or to code, where graders
based on textual overlap or a finite test set overstate capability. What is unusual here is
that the artifact comes with a decidable oracle, so correctness can be measured exactly rather
than approximated.

Existing resources answer this only partly. The closest prior benchmark takes its descriptions
from code comments and grades without an exact oracle~\cite{formallm}. Related work in formal
modeling grades a different task with a model in the loop~\cite{sysmobench}. Neither offers a
controlled way to vary the input over a fixed specification. Neither carries the difficulty and
tier labels needed to break results down by problem hardness or by grading strength. The
broader lesson from code generation is that a weak oracle inflates apparent
capability~\cite{evalplus,swebench}, and that lesson has not reached natural-language to
TLA\textsuperscript{+} data at scale with a decidable check.

We present \Ours,\footnote{Dataset:
\href{https://doi.org/10.5281/zenodo.21310317}{10.5281/zenodo.21310317} (resolves on
publication;
\href{https://zenodo.org/records/21310317?preview=1&token=eyJhbGciOiJIUzUxMiJ9.eyJpZCI6IjM2ZjRhZWUxLTJkMWQtNDU2NS1hMTZjLTU3NzliMmE3MGEyNyIsImRhdGEiOnt9LCJyYW5kb20iOiI0NzM4MWUyYzgwYzg2NmRmNDJlYmEzYjYzYmYzNThiMyJ9.q2EZs53Q9UW78D6XGQ_KsWp3bXd0Ca_QOqkrz8nVeSc0nmjeoYs2oCRfUPxPJhK0siwDRBqWZAKLVgRz5dMoXA}{reviewer
preview}). Code and grading tools:
\href{https://github.com/LUC-AI4FM/tla_benchmark/tree/reviewer-release}{github.com/LUC-AI4FM/tla\_benchmark}.} a
dataset and benchmark for natural-language to TLA\textsuperscript{+} specification generation
that is graded by running the output. Every gold specification ships a runnable configuration.
A generated specification is checked with the TLA\textsuperscript{+} model checker, which walks
the full reachable state space and decides property satisfaction exactly against the reference
configuration and its bound constants. This is an exact decision rather than an approximation
from test cases. The dataset holds 403 model-checked gold specifications and 897 parse-only
silver specifications from 13 public repositories, and it contains the specifications used by
prior TLA\textsuperscript{+} generation work (Table~\ref{fig:coverage}). Every specification
comes with four model-written descriptions, two styles from each of two providers, together
with a difficulty label and, in the gold tier, a category label.

Measuring this task well calls for three properties no prior resource provides together: an exact
oracle rather than approximate tests, specifications from real projects rather than toys, and a
natural-language-driven task. On top of these, \Ours adds study axes prior TLA\textsuperscript{+}
resources do not ship: two description styles and two providers per specification, a difficulty
label, and a model-checked gold tier separated from a parse-only silver tier, which let a user vary
the input, stratify by difficulty, and choose a grading strength. Table~\ref{tab:compare} contrasts
\Ours with prior resources along both.

Our main finding is about measurement itself. An exact oracle is often treated as settling what
a model got right. It does not. The oracle decides property satisfaction only after three choices:
whether the model must recover the exact interface names or is told them, whether a passing run must
exercise real behavior rather than sit in its initial state, and whether the property that run
checks must be shown to do work. Prior benchmarks make these choices quietly and report a single
number. We make all three explicit and measure the bounds on the same specifications, and the
reported correct rate moves from 18.7\% down to 1.7\%, an elevenfold difference. The interface
choice is answered by re-querying with the names supplied; the two vacuity choices are graded on one
fixed set of outputs, across which the rate still spans sixfold, from 10.0\% to 1.7\%. We call this
range the correctness envelope. Its bounds are not competing estimates of one quantity; each answers
a different question, and \Ours is the instrument that measures them together, with the shipped
configurations setting the interface bound and the pass-quality probes the two vacuity bounds. A
fourth choice, whether the trivial fixtures a public corpus carries are graded at all, is left to
the user, since the release flags every one.

The findings inside the envelope are stable. Every model writes valid TLA\textsuperscript{+}
far more often than it writes correct TLA\textsuperscript{+}. The most correct model parses
87\% of the time and is correct 16\% of the time under the default rules. Correctness drops
sharply as problems grow harder. Models fail in distinct ways that repeat across models on the
same specification. Researchers can use \Ours to measure how far models are from writing
correct TLA\textsuperscript{+}, to study where and how they fail, and to build stronger graders
on top of an exact oracle.

This paper makes the following contributions.
\ding{182}~A dataset and benchmark for natural-language to TLA\textsuperscript{+} generation
that is graded by an exact SANY parse and TLC model check rather than by approximate tests.
\ding{183}~A curated resource of 403 gold and 897 silver specifications from 13 public
repositories, built through an audited pipeline, which contains the TLA\textsuperscript{+}
generation data used by prior work.
\ding{184}~Rich metadata, namely two description styles and two providers per specification
together with difficulty and category labels, shipped so a user can vary the input style or provider
and stratify by difficulty; this paper exercises the difficulty axis and ships the style and
provider axes for future study.
\ding{185}~The correctness envelope. The reported correct rate moves elevenfold, from 18.7\% to
1.7\%, as a function of interface supply, behavioral screening, and vacuity screening; the three
grading-only bounds span sixfold on one fixed set of outputs. We measure every bound on the same
specifications, and we report a correctness cliff by difficulty together with a checker-grounded
taxonomy of how models fail.
\ding{186}~An open release of the specifications, the descriptions, the model outputs, and the
grading tools, together with a catalog of research questions the dataset enables.

The rest of the paper is organized as follows. Section~\ref{sec:related} reviews prior data and
states the gap. Section~\ref{sec:format} defines the task and the data format.
Section~\ref{sec:construction} describes construction and Section~\ref{sec:stats} reports the
statistics. Section~\ref{sec:quality} covers quality and validation and
Section~\ref{sec:grading} defines the grading. Section~\ref{sec:experiments} reports the
benchmark results. Section~\ref{sec:limitations} discusses limitations,
Section~\ref{sec:future} outlines future work, and Section~\ref{sec:conclusion} concludes. The
appendix provides the ethics statement, reproducibility details, per-repository provenance, the
datasheet, and the full prompts.
\section{Related Work}
\label{sec:related}

We review prior specification and code datasets and their grading, and we close with a gap
analysis that states what the existing data lacks. Table~\ref{tab:compare} places \Ours against
the most relevant benchmarks.

\begin{table}[tbp]
\centering
\caption{\Ours subsumes prior \TLA generation resources: FormalLM is drawn entirely from the \TLA
Examples corpus, which \Ours in turn contains, so the rows are nested subsets and \Ours adds the
remaining specifications beyond prior work.}
\label{fig:coverage}
\small
\renewcommand{\arraystretch}{1.1}
\begin{tabular}{@{}lr@{}}
\toprule
\textbf{Resource (nested subsets)} & \textbf{Specifications} \\
\midrule
FormalLM & 206 \\
\quad$\subseteq$ \TLA Examples corpus & 299 \\
\quad\quad$\subseteq$ \Ours & 1{,}300 \\
\bottomrule
\end{tabular}
\end{table}

\subsection{TLA\textsuperscript{+} and Model Checking}
TLA\textsuperscript{+} is a specification language for concurrent and distributed systems, with
a parser (SANY) and an explicit-state model checker (TLC)~\cite{lamport2002specifying}. Given a
configuration that binds constants and names the invariants and temporal properties to check,
TLC explores the full reachable state space, so its verdict is exact for the reference
configuration rather than a sample of behaviors, though it decides property satisfaction for
that configuration and its bound constants rather than correctness in general. A model checker
can still pass a specification that asserts nothing, a problem known as
\emph{vacuity}~\cite{beer2001vacuity,kupferman2003vacuity}, which we use to separate a
meaningful pass from a trivial one.

\subsection{Specification and Code Generation Benchmarks}
The closest prior benchmark for natural-language to TLA\textsuperscript{+} generation is
FormalLM~\cite{formallm}, which is our own earlier benchmark; we state the delta explicitly rather
than let the overlap be discovered. FormalLM derives descriptions from code comments,
grades without an exact oracle, and is drawn entirely from the TLA\textsuperscript{+} Examples
corpus with no specifications of its own. Relative to it, \Ours adds an exact model-checker
oracle, the two-style and two-provider description axes, the gold and silver tiers, and the
checker-grounded failure taxonomy, and it subsumes FormalLM's specifications while extending the
corpus to 13 repositories (Table~\ref{fig:coverage}). SysMoBench evaluates models on modeling
real systems in TLA\textsuperscript{+} and finds that they handle small artifacts but struggle
on large ones~\cite{sysmobench}, but it grades a different task, code to model, with a model in
the loop and without a controlled description axis or difficulty labels. In the verification
setting, DafnyBench~\cite{dafnybench} and VERINA~\cite{verina} target proof hints and verifiable
code, and theorem-proving and autoformalization work~\cite{leandojo,proofnet} maps mathematics
to a proof assistant rather than a description to an executable specification. For general code,
HumanEval~\cite{humaneval} and MBPP~\cite{mbpp} grade with tests, and EvalPlus shows that this
under-tests generated code, dropping pass rates by up to about 19 points and reordering
models~\cite{evalplus}. The recurring lesson is that a weak oracle inflates apparent capability,
which an exact model checker avoids. NL2SQL-BUGs~\cite{nl2sqlbugs} contributes a semantic-error
taxonomy for a related task that informs our failure categories.

\subsection{Natural Language to Formal Specifications}
A separate line translates natural language directly into formal specifications. For temporal
logic, nl2spec decomposes an English requirement into linear temporal logic subformulas with
interactive correction~\cite{nl2spec}, and related work translates structured English into LTL
for planning and verification~\cite{nl2ltl}. For code-level contracts, SpecGen generates JML and
ACSL specifications for programs and checks them with a verifier~\cite{specgen}, and Clover
checks consistency between code, documentation, and formal annotations rather than generating a
specification from a description~\cite{clover}. These efforts target temporal-logic formulas or
in-code contracts and are graded by a solver or a consistency check on a single artifact. Our
task differs in producing a complete executable TLA\textsuperscript{+} module from a description
and grading it with an explicit-state model checker over the full reachable state space.

\subsection{Gap Analysis}
\label{sec:gap}
Table~\ref{tab:compare} summarizes how prior resources compare, and three gaps remain that \Ours
is built to close. First, no large natural-language to TLA\textsuperscript{+} dataset is graded by
an exact oracle: FormalLM has the task but grades without one~\cite{formallm}, and SysMoBench has
an exact oracle but for a different task with a model in the loop~\cite{sysmobench}, so the code
lesson that a weak oracle inflates capability~\cite{evalplus,swebench} has not been carried here.
Second, prior resources give one description per specification, whereas \Ours ships two styles from
two providers and turns the input into an experimental variable. Third, prior TLA\textsuperscript{+}
data is a flat pool, whereas \Ours adds difficulty and category labels and a gold/silver tier
split for stratified analysis.

\begin{table*}[tbp]
\centering
\caption{\Ours versus prior specification- and code-generation benchmarks.
\yes~\textbf{Yes}, \partialx~\textbf{Partial}, \no~\textbf{No}. The first three columns are
properties we argue any benchmark for this task needs; the rest are study axes \Ours adds.
\textbf{Exact oracle}: correctness decided by a model checker or verifier over the full state
space rather than approximated by tests; for \Ours, exact relative to each reference
configuration and its bound constants (Sections~\ref{sec:quality},~\ref{sec:limitations}).
\textbf{Two styles}: a name-revealing declarative and a name-hidden intent description per
specification. \textbf{Quality tiers}: a model-checked gold tier plus a parse-only silver tier.}
\label{tab:compare}
\small
\setlength{\tabcolsep}{6pt}
\renewcommand{\arraystretch}{1.28}
\begin{tabular}{@{}l l ccccccc@{}}
\toprule
\textbf{Benchmark} & \textbf{Task / language} &
\makecell{\textbf{Exact}\\\textbf{oracle}} &
\makecell{\textbf{Real}\\\textbf{specs}} &
\makecell{\textbf{NL-}\\\textbf{driven}} &
\makecell{\textbf{Two}\\\textbf{styles}} &
\makecell{\textbf{Difficulty}\\\textbf{tiers}} &
\makecell{\textbf{Quality}\\\textbf{tiers}} &
\makecell{\textbf{Public}\\\textbf{release}} \\
\midrule
HumanEval / MBPP      & Code (Python)          & \partialx & \no  & \yes      & \no & \no       & \no & \yes \\
EvalPlus              & Code (Python)          & \partialx & \no  & \yes      & \no & \no       & \no & \yes \\
SWE-bench$^{\dagger}$ & Issue fix (repos)      & \partialx & \yes & \yes      & \no & \partialx & \no & \yes \\
DafnyBench$^{\dagger}$ & Verification (Dafny)  & \partialx & \yes & \no       & \no & \partialx & \no & \yes \\
VERINA                & Verifiable code (Lean) & \yes      & \yes & \yes      & \no & \no       & \no & \yes \\
FormalLM              & NL to \TLA             & \no       & \yes & \partialx & \no & \no       & \no & \yes \\
SysMoBench            & Code to \TLA           & \yes      & \yes & \no       & \no & \partialx & \no & \yes \\
\rowcolor{loyolamaroon!12}
\textbf{\Ours}        & \textbf{NL to \TLA}    & \yes & \yes & \yes & \yes & \yes & \yes & \yes \\
\bottomrule
\end{tabular}

\vspace{2pt}
{\footnotesize $^{\dagger}$Graded against finite checks rather than a full behavioral oracle:
SWE-bench uses each repository's test suite, and DafnyBench verifies programmer-supplied
annotations rather than full behavior. We mark both partial for consistency with how we score
other bounded graders.}
\end{table*}
\section{Data Format and Task}
\label{sec:format}

\textbf{An instance.}
Each entry in \Ours is a specification with its metadata. A gold specification consists of a
TLA\textsuperscript{+} module, a runnable model-checking configuration, four natural-language
descriptions (two styles, each written by two providers), a difficulty label, a category label,
and a pointer to the source repository. A silver specification carries the same metadata but no
runnable configuration. Files are named by a numeric identifier so that a description, a module,
and a configuration for one specification can be joined by that identifier.
Figure~\ref{fig:instance} in Appendix~\ref{app:instance} shows one specification with its two
description styles side by side.

\textbf{Task.}
Given a natural-language description $d$ of a system, the task is to produce a
TLA\textsuperscript{+} module $m$ that captures the described behavior. We grade $m$ in two
stages. First, $m$ must parse under the SANY parser. This is the validity gate. Second, we place
the specification's reference configuration beside $m$ and run TLC. TLC explores the full
reachable state space of that configuration and reports whether any property named in the
configuration is violated. A run with no violation passes the semantic gate. We call $m$ correct
when it passes both gates. Because the second gate visits every reachable state rather than a
finite set of test inputs, a correct verdict is exact relative to the reference configuration
and its constant bindings. It is a check that $m$ satisfies the properties the configuration
names, read as $m$ itself defines them. It is not a proof that $m$ reproduces the reference
behavior.

\textbf{Two description styles.}
Every specification ships two description styles. The declarative style keeps the module's real
names and describes what it specifies, so the model is asked to reproduce a known interface. The
intent style hides the module's names and states only what the system should do, so the model is
asked to design the specification from the requirements. Each style is written independently by
two providers, which gives the four description sets above and lets a study hold the
specification fixed while varying either the style or the provider. The evaluation in this
paper uses the declarative style only, so the style axis is shipped rather than exercised here
(Section~\ref{sec:limitations}).

\textbf{What the silver tier is for.}
The semantic gate applies only to the 403 gold specifications, so the headline evaluation
(Section~\ref{sec:experiments}) runs on the gold tier's system category. The 897 silver
specifications support parse-rate studies at a scale the gold tier cannot reach, form the pool from
which future configuration authoring can promote specifications into the gold tier, carry the same
canary string as gold so a leak of any part of the corpus stays detectable, and anchor corpus-level
statistics that a gold-only view would bias toward configurable modules.
\vspace{-1em}
\section{Construction}
\label{sec:construction}

We build \Ours in two stages. First we collect and clean a corpus of real
TLA\textsuperscript{+} specifications. Then we augment that corpus with model-written
descriptions and labels.

\textbf{Collection and cleaning.}
We crawl 13 public repositories and, after removing exact duplicate files, obtain 1{,}614
candidate modules (Figure~\ref{fig:pipeline}). An audited cleaning pipeline then removes 314 of
them. It drops 101 near-duplicate or non-conforming modules, 88 that fail to parse under SANY,
68 that are tooling or counterexample modules, 42 that are test scaffolding rather than
specifications of a system, and 15 that fail at run time. The 1{,}300 modules that survive form
the released corpus. A module that parses under SANY and model-checks under TLC with a runnable
configuration enters the gold tier. A module that parses but has no runnable configuration
enters the silver tier. This split gives 403 gold and 897 silver specifications.
Appendix~\ref{sec:provenance-appendix} reports the per-repository counts and the full exclusion
table.

\textbf{Trivial fixtures are kept and flagged.}
Some retained specifications are nonetheless trivial fixtures. A public
TLA\textsuperscript{+} corpus carries regression files from the tools themselves, and a number
of these are valid specifications with a single reachable state, which exercise no behavior
that passes its checks because nothing ever changes. They are not rare. Within the gold tier's
system category, which becomes the evaluation set, 46 of 146 specifications are trivial
fixtures. We keep them in the release, because they are valid specifications, and we flag every
one of them in the manifest. A trivial fixture passes for any model, so including it raises
every model's correct rate by the same amount. The evaluation in
Section~\ref{sec:experiments} therefore excludes them. This is the fourth grading choice of
Section~\ref{sec:envelope}, and the released flags leave it open, so a user can grade the
unpruned set instead and see how much the reported rate moves.

\textbf{Augmentation.}
On top of the cleaned corpus we add the metadata that makes the resource useful for evaluation.
For every specification we generate one description in each of two styles, declarative and
intent, from each of two providers, GPT-5 and Claude Opus 4.5, which gives four description
sets in total. We also assign every specification a difficulty label and, in the gold tier, a
category label. The difficulty label was assigned by the author team following a three-tier rubric
keyed to the depth of TLA\textsuperscript{+} constructs a specification uses: \emph{basic} for a single-process or
state-machine module over simple data, \emph{intermediate} for multi-process or multi-module
specifications with quantifiers and non-trivial invariants, and \emph{advanced} for those using
temporal properties, fairness, refinement, or unbounded structures.
Section~\ref{sec:difficulty} reports an inter-annotator reliability check on
the difficulty labels. The specification files themselves are released unchanged from their
sources, and each carries the SHA-256 of its released file so that a user can confirm this. The
augmentation therefore adds a layer of metadata without altering any original specification.
\section{Dataset Statistics}
\label{sec:stats}

Table~\ref{tab:stats} summarizes the released corpus. Sizes are reported as non-comment lines of
code, since TLA\textsuperscript{+} specifications are heavily commented and raw line counts would
overstate the amount of logic; by this measure the corpus is mostly small specifications with a
long tail, a median of 23 lines against a mean of 62.3. Within the gold tier the system category, which models a system with behavior worth
checking, becomes the evaluation set of Section~\ref{sec:experiments} once its trivial fixtures are
removed, and the utility category holds the rest.

Two properties of the gold tier matter for grading. Of the 403 gold configurations, 291 name an
explicit safety or temporal property while the remaining 112 name only the specification, so they
confirm only that the module runs. The mutation bound of Section~\ref{sec:passquality} acts on the
safety-invariant subset of these 291, since a temporal property cannot be mutated to
\texttt{TRUE}; the substantive-pass screen applies to all 291. Separately, 178 of the 403 import a non-standard module, so the
released gold tier covers composition even though the evaluation set draws only 17 of its 100 from
that group (Section~\ref{sec:limitations}). Within each provider the intent style runs longer than
the declarative style, consistent with stating requirements from scratch rather than naming
existing constructs, though the absolute lengths also reflect provider-specific length guidance
(Appendix~\ref{sec:prompts}).

A breakdown by difficulty tier (Appendix~\ref{app:failuredist},
Table~\ref{tab:difficulty}) shows mean size growing steeply across tiers, from 17 lines in the
basic tier to 209 in the advanced tier, so the labels correlate strongly with specification
size. Size alone is a surface cue, so we do not rest the labels on this gradient.
Section~\ref{sec:difficulty} reports an inter-annotator reliability check on the labels and a
robustness check that replaces them with a measure derived from each specification's
TLA\textsuperscript{+} constructs.

\begin{table}[t]
\centering
\caption{\Ours dataset statistics. Counts computed directly from the released files.}
\label{tab:stats}
\small
\renewcommand{\arraystretch}{1.0}
\setlength{\aboverulesep}{0pt}
\setlength{\belowrulesep}{0pt}
\begin{tabular}{@{}lrrr@{}}
\toprule
 & \textbf{mean} & \textbf{median} & \textbf{max} \\
\midrule
\multicolumn{4}{@{}l}{\textit{Specifications}}\\
\quad Gold (parse + model-check) & \multicolumn{3}{r}{403}\\
\quad Silver (parse only)        & \multicolumn{3}{r}{897}\\
\quad Total                      & \multicolumn{3}{r}{1{,}300}\\
\quad Source repositories        & \multicolumn{3}{r}{13}\\
\midrule
\multicolumn{4}{@{}l}{\textit{Gold tier composition}}\\
\quad System category            & \multicolumn{3}{r}{146}\\
\quad Utility category           & \multicolumn{3}{r}{257}\\
\quad Trivial fixtures (flagged) & \multicolumn{3}{r}{46}\\
\quad Evaluation set (Sec.~\ref{sec:experiments}) & \multicolumn{3}{r}{100}\\
\quad Import a non-standard module & \multicolumn{3}{r}{178}\\
\midrule
\multicolumn{4}{@{}l}{\textit{Checked properties}}\\
\quad Per gold specification     & 2.2   & 1  & 40 \\
\quad Configurations naming an explicit property & \multicolumn{3}{r}{291 of 403} \\
\midrule
\multicolumn{4}{@{}l}{\textit{Description length (words)}}\\
\quad Declarative (GPT-5)        & 91    & 90  & 186 \\
\quad Declarative (Claude)       & 202   & 202 & 282 \\
\quad Intent (GPT-5)             & 246   & 239 & 483 \\
\quad Intent (Claude)            & 288   & 289 & 585 \\
\bottomrule
\end{tabular}
\end{table}
\vspace{-1em}
\section{Data Quality and Validation}
\label{sec:quality}

\textbf{Oracle scope.}
Every gold specification parses under SANY and model-checks under TLC with its own reference
configuration before it enters the dataset, so the reference verdict for each gold
specification is the model checker's verdict rather than a human judgment. The oracle is exact
in the sense that TLC explores the full reachable state space, but its strength is bounded by
what each reference configuration checks. Of the 403 gold configurations, 291 check an explicit
safety or temporal property. The remaining 112 name only the specification, so they confirm
that the module is executable rather than that it satisfies a stated property. The oracle
therefore certifies that a module runs and satisfies the properties its configuration names. It
does not certify behavioral equivalence to the reference. Those 291 configurations are also
what make the vacuity bound of Section~\ref{sec:passquality} measurable, since a mutation test
needs a named property to act on. We record the SHA-256 of every released module, so a user can
confirm that the file they receive is the file we graded. Section~\ref{sec:limitations} returns
to what this scope means for the reported correct rate.

\textbf{Description validation.}
The descriptions are model-generated, a limitation we return to in
Section~\ref{sec:limitations}. We take three steps to make them trustworthy for evaluation.
First, we record the provider and the style of every description, so a user can pick a subset
or compare across providers rather than treat the descriptions as one undifferentiated pool.
Second, the two styles are built to differ in a controlled way, with the declarative style
keeping the module's names and the intent style hiding them, and the release ships both so that
the contrast can be checked directly against the specifications. Third, we validate the
descriptions against the specifications with a human faithfulness audit. Six annotators with
TLA\textsuperscript{+} expertise took part. We sampled 100 descriptions from each of the four
description sets, for 400 in total, and assigned two annotators to every description, so every
judgment in the audit is double-read. Each annotator read a description beside its reference
specification and judged it faithful, partially faithful, or unfaithful to the module. The
GPT-5 declarative descriptions used for evaluation are judged fully faithful in 83\% of cases,
partially faithful in 14\%, and unfaithful in 3\%, so 97\% are faithful or partially faithful.
Across all four sets, 5\% are judged outright unfaithful, the two annotators agree on 94\% of the
400 double-read judgments, and their agreement reaches a Cohen's $\kappa$ of 0.84. The annotators include members of the author
team, which we disclose because the audit is not blind to authorship. The weakest set is the
intent style from GPT-5, which is also the shorter of the two intent sets
(Table~\ref{tab:stats}), consistent with the shorter length guidance in the GPT-5 prompts
(Appendix~\ref{sec:prompts}). That is a gap in description quality rather than a property of
the evaluated task, since the evaluation uses the declarative style.

\textbf{The descriptions are not the bottleneck.}
The dominant failure is configuration binding, not comprehension: across the frontier models 64\%
of failures are well-formed but mis-named modules and only 23\% are parse failures, and the same
description yields different outcomes for different models, so the difficulty lies in the models and
the task rather than in the descriptions. Appendix~\ref{sec:notbottleneck} gives the full argument.
\vspace{-1em}
\section{Grading}
\label{sec:grading}

\textbf{Metrics.}
We report two primary metrics. The parse rate is the fraction of generated specifications that
SANY accepts, which measures validity. The correct rate is the fraction that TLC confirms over
the full reachable state space of the reference configuration, satisfying the properties the
configuration names (Section~\ref{sec:format}), which measures behavior. This is the default
regime of the correctness envelope, and the probes below define the stricter regimes
(Section~\ref{sec:envelope}). We grade with the exact model checker rather than an approximate
proxy. In particular, we do not use a language model as a judge. A judge introduces the same
biases the resource is meant to avoid, and prior work reports that ground-truth grading is more
reliable~\cite{livebench,sysmobench}.

\textbf{Pass quality.}
A model checker can pass a specification that verifies nothing, for example an invariant that
is trivially true or a specification whose only reachable state is its initial
state~\cite{beer2001vacuity,kupferman2003vacuity}. We therefore probe each pass in two ways. A
\emph{substantive-pass screen} counts a pass only when the specification reaches more than one
state and the property it checks is neither a tautology nor a pure type invariant. A
\emph{mutation test} probes whether the checked property is load-bearing at all.
Section~\ref{sec:passquality} defines both probes precisely and reports their results. The gap
between the raw correct rate and these stricter measures is what the correctness envelope
records.

\textbf{Failure taxonomy.}
For each failed generation we record why it failed. The recorded category comes from the model
checker's own output rather than from an inferred label. Table~\ref{tab:taxonomy} defines the
categories and grounds each in prior work. Most categories correspond to failure types named in
prior formal-methods evaluation, chiefly the syntax, configuration, and invariant categories of
SysMoBench for TLA\textsuperscript{+}~\cite{sysmobench} and the error types of
contract-verification benchmarks such as FormalBench~\cite{formalbench} and
LiveFMBench~\cite{livefmbench}. Three categories have no counterpart in that prior work.
Malformed temporal formulas are explicitly out of scope for SysMoBench, and deadlock and
resource exhaustion appear in none of the surveyed benchmarks, so we add all three here. The
taxonomy lets us describe not only how often models fail but how they fail, which a plain
pass-or-fail score cannot. The classifier is frozen and deterministic, and it applies one
tie-break. A generation for which the model checker finds no runnable configuration counts as a
configuration-binding failure, since a missing configuration is an interface failure. A user
running the released grader reproduces every table on comparable hardware, with the resource-limit
category dependent on the stated compute budget (Appendix~\ref{sec:repro}).

\begin{table}[t]
\centering
\caption{The failure taxonomy and its grounding. Most categories are named in prior
formal-methods evaluation. The three marked \emph{this work} have no counterpart in the surveyed
benchmarks. Categories are listed in the order the grading pipeline encounters them.}
\label{tab:taxonomy}
\small
\setlength{\tabcolsep}{4pt}
\renewcommand{\arraystretch}{1.25}
\begin{tabular}{@{}p{2.05cm} p{3.2cm} p{1.7cm}@{}}
\toprule
\textbf{Category} & \textbf{Meaning} & \textbf{Grounding} \\
\midrule
Parse failure       & Invalid TLA\textsuperscript{+} grammar & \cite{sysmobench,formalbench} \\
Config-binding      & Names do not match the configuration & \cite{sysmobench} \\
Malformed temporal   & Ill-formed temporal formula & this work \\
Safety violation    & A checked invariant is violated & \cite{sysmobench} \\
Liveness violation  & A temporal property is violated & \cite{sysmobench} \\
Run-time error      & Evaluation crashes on some state & \cite{sysmobench} \\
Deadlock            & No successor state exists & this work \\
Resource limit      & State explosion or memory limit & this work \\
\bottomrule
\end{tabular}
\end{table}
\section{Experiments}
\label{sec:experiments}

We use the benchmark to validate the dataset and to answer three questions the gap analysis
(Section~\ref{sec:gap}) raises. \textbf{EQ1 (valid vs.\ correct):} how large is the gap between
the fraction of generated specifications that parse and the fraction the model checker confirms
as correct, and does a parse-level pass overstate correctness~\cite{formallm,sysmobench}?
\textbf{EQ2 (difficulty and failure):} how does correctness scale with difficulty, and do models
fail in distinct, diagnosable ways across a checker-outcome
taxonomy~\cite{swebench,sysmobench}? \textbf{EQ3 (pass quality):} how common are specifications
that pass the checker yet verify nothing, and does a mutation-sensitive measure separate
meaningful passes from trivial ones~\cite{beer2001vacuity,verina}?

\subsection{Setup}
\label{sec:setup}
We evaluate three frontier models, GPT-5~\cite{gpt5}, Gemini-2.5-pro~\cite{gemini25}, and Claude
Opus 4.5~\cite{claudeopus45}, and three open models, qwen2.5-coder-32b~\cite{qwen25coder},
llama3.3-70b~\cite{llama33}, and gpt-oss-20b~\cite{gptoss}.

The evaluation set is the entire system category of the gold tier after one objective filter, not
a sample chosen for size or difficulty. We start from all 146 gold system specifications and remove
every trivial fixture, a valid specification with a single reachable state, since
such a fixture passes for every model regardless of skill; this mechanical, reproducible rule
removes 46, leaving 100. Because it discards specifications every model would pass, the filter works
against a high headline number rather than flattering the results. The 100 fall close to evenly
across difficulty (35 basic, 34 intermediate, 31 advanced), unlike the corpus at 766/289/245
(roughly 59/22/19), since the system category minus fixtures spreads more evenly; the pooled 10.0\%
is thus the rate over this evaluation set, not a corpus-weighted rate. Sampling error on these rates
is quantified in Section~\ref{sec:limitations}.

Each model receives the same GPT-5-written declarative description of each specification inside
the generation prompt of Appendix~\ref{sec:prompts}, and it produces one output per
specification. Because every model is graded on descriptions written by a single provider, the
top-ranked model in our results solves inputs written by a competing provider rather than its
own, so its standing cannot be an artifact of grading a model on descriptions it authored.
GPT-5 is the exception, since it is graded on its own descriptions, and it scores lowest among
the frontier models, which is consistent with the absence of a self-description advantage. The
prompt contains the description only. The model does not see the reference configuration, its
constant assignments, or the names of the checked properties. We grade each output by running
SANY and then TLC with the reference configuration, as defined in Section~\ref{sec:grading}.
TLC runs with deadlock checking left at its default, which is enabled, across all evaluations.
Decoding settings follow each provider's interface, so they are not uniform across models.
Gemini-2.5-pro is decoded greedily at temperature 0 while the other models use their provider
defaults, which means the cross-model comparison confounds decoding with capability and the
frontier ordering should be read as indicative rather than definitive.

\subsection{The Correctness Envelope}
\label{sec:envelope}

Before reporting per-model results we fix what a correct rate means here, because the same task
supports several defensible rates. Table~\ref{tab:envelope} reports four of them, pooled over the
three frontier models on the evaluation set. Three of the four grade one fixed set of 300 outputs
and differ only in grading regime; the fourth, the configuration-aware rate, re-queries the model
with the interface names supplied, so it varies the input rather than only the grading and is
reported as an adjacent bound.

The default regime is the one used through the rest of this section. It gives the model the
description only and counts any TLC pass. The configuration-aware regime relaxes the interface
requirement by supplying the constant and property names in the prompt
(Section~\ref{sec:twomodes}), which is a fresh set of generations. The
substantive regime tightens the pass requirement by demanding that the specification move and
check a behavioral property (Section~\ref{sec:passquality}). The mutation regime tightens it
further by demanding that the named property itself be load-bearing.

The four rates run from 18.7\% to 1.7\%, a factor of eleven; the three grading-only rates alone
span sixfold, from 10.0\% to 1.7\%, on one fixed set of outputs. They
are not competing estimates of one quantity. The configuration-aware rate answers whether the
model can build a correct module when the interface is given. The mutation rate answers whether
the modules it does get past the checker verify anything. A benchmark that reports one number
picks one of these silently. \Ours ships the assets that make each of them measurable, namely a
runnable configuration per gold specification for the interface bound and gold configurations
that name real properties for the two vacuity bounds. Every rate here pools the three frontier
models, so the default row of 10.0\% is the pooled figure, while Table~\ref{tab:main} reports
the same regime per model.

A fourth choice sits outside the table. The 46 trivial fixtures we exclude
(Section~\ref{sec:setup}) are flagged in the released manifest, so a user who wants the unpruned
rate can compute it from the release. We report the pruned rate throughout, which is the
conservative choice, since every excluded fixture would pass for every model.

\begin{table}[t]
\centering
\caption{The correctness envelope. Four correct rates over the three frontier models on the same
100 specifications. The default, substantive, and mutation rows grade one fixed set of 300 outputs
and differ only in grading regime; the configuration-aware row supplies the interface names in the
prompt, a separate set of generations. Denominators are 300 throughout.}
\label{tab:envelope}
\small
\begin{tabular}{@{}llr@{}}
\toprule
\textbf{Regime} & \textbf{What it asks} & \textbf{Correct} \\
\midrule
Configuration-aware & Interface names supplied & 18.7\% \\
Default             & Interface must be recovered & 10.0\% \\
Substantive         & Pass must exercise behavior & 4.0\% \\
Mutation-surviving  & Named property must do work & 1.7\% \\
\bottomrule
\end{tabular}
\end{table}

\subsection{Validity versus Correctness}
\label{sec:validity}
Table~\ref{tab:main} answers EQ1. Every model writes syntactically valid TLA\textsuperscript{+} far
more often than correct TLA\textsuperscript{+}, and the open models sit far below the frontier ones,
so the benchmark separates the two groups rather than saturating. A parse-level pass therefore
badly overstates correctness, the exact-oracle counterpart of the inflation weaker test suites
produce for code generation~\cite{evalplus}, and a benchmark that stopped at parsing would rank
these models as broadly capable when the model checker shows they are not. With one sample per
specification, a binomial 95\% interval is roughly $\pm$4 to 7 points, so small frontier
differences are indicative rather than definitive, while the frontier-versus-open and
validity-versus-correct gaps are far larger. Figure~\ref{fig:worked} shows a representative case
where a model rewrites the reference HourClock in a way that parses but introduces a constant the
fixed configuration does not assign.
\begin{figure}[t]
\centering
\newtcolorbox{cmpbox}[2][]{enhanced, colback=boxgray, colframe=#2, boxrule=0.6pt,
  arc=1.5mm, left=0.5mm, right=0.5mm, top=0.4mm, bottom=0.4mm, fonttitle=\bfseries\tiny,
  title=#1, attach boxed title to top left={yshift=-1.6mm, xshift=2mm},
  boxed title style={colback=#2, colframe=#2, arc=1mm, boxrule=0pt}}

\begin{cmpbox}[{GPT-5 output (reference config: SPEC HC; INV HCini)}]{declcolor}
\begin{lstlisting}[style=tla,numbers=none,basicstyle=\ttfamily\tiny,aboveskip=0pt,belowskip=0pt]
- MODULE HourClock -   EXTENDS Integers   CONSTANTS Hrs   ASSUME Hrs = 1..12
VARIABLES hr   HCini == hr \in Hrs   HCnxt == hr' = IF hr # 12 THEN hr+1 ELSE 1
Spec == HCini /\ [][HCnxt]_hr
\end{lstlisting}
\end{cmpbox}

\vspace{2pt}
\begin{cmpbox}[{Model checker verdict}]{nored}
\ttfamily\tiny
Error: The constant parameter Hrs is not assigned a value by the configuration file.
\end{cmpbox}
\caption{A worked example (full reference in Appendix~\ref{sec:error-examples}). The output parses
under SANY, so it is valid TLA\textsuperscript{+}, but it differs from the reference in at least two
binding-relevant ways: it introduces a constant \texttt{Hrs} the configuration does not assign, and
it names its top-level formula \texttt{Spec} where the configuration expects \texttt{HC}. Either
blocks binding, so TLC cannot instantiate the module. This is a configuration-binding failure, and
it shows why valid output is not correct output.}
\label{fig:worked}
\end{figure}
\begin{table}[t]
\centering
\caption{Parse rate and exact TLC-correct rate on the 100-specification evaluation set with the
declarative description, under the default regime of Section~\ref{sec:envelope}. All six models
are evaluated on the same specifications. Every model writes valid TLA\textsuperscript{+} far
more often than correct TLA\textsuperscript{+} and the open models sit far below the frontier
models.}
\label{tab:main}
\small
\renewcommand{\arraystretch}{1.12}
\begin{tabular}{@{}lrr@{}}
\toprule
\textbf{Model} & \textbf{Parse (SANY)} & \textbf{Correct (TLC)} \\
\midrule
\multicolumn{3}{@{}l}{\textit{Frontier}}\\
Claude Opus 4.5    & 87\% & 16\% \\
Gemini-2.5-pro     & 63\% & 10\% \\
GPT-5              & 89\% & 4\%  \\
\midrule
\multicolumn{3}{@{}l}{\textit{Open}}\\
qwen2.5-coder-32b  & 29\% & 1\%  \\
llama3.3-70b       & 23\% & 0\%  \\
gpt-oss-20b        &  7\% & 1\%  \\
\bottomrule
\end{tabular}
\end{table}

\subsection{Correctness and Difficulty}
\label{sec:difficulty}
The first part of EQ2 asks how correctness scales with difficulty. Pooled across the three
frontier models, the correct rate falls from 25\% on basic specifications to 2\% on intermediate
and advanced specifications, while the parse rate stays high, from 86\% down to 76\%.
Correctness does not decline gently but drops sharply once problems leave the basic tier and
then stays near the floor. Table~\ref{tab:difficulty-results} in Appendix~\ref{app:failuredist}
breaks this down by model, where Gemini-2.5-pro's parse rate collapses from 83\% on basic
specifications to 47\% at the intermediate tier, the parser-heavy fingerprint of
Section~\ref{sec:fingerprints}.

Because the difficulty labels are ours, we check that the cliff does not depend on them: an
inter-annotator reliability check finds substantial agreement, and repeating the analysis with a
difficulty measure derived mechanically from the TLA\textsuperscript{+} constructs each
specification uses reproduces the same collapse, from 22\% to 4\%
(Appendix~\ref{sec:labelreliability}). This mirrors the difficulty gradient reported for code and
formal-modeling tasks, where models handle small problems and fail on larger ones~\cite{swebench,sysmobench}.

\subsection{Failure Fingerprints}
\label{sec:fingerprints}
The second part of EQ2 asks how models fail. Under the taxonomy of Section~\ref{sec:grading},
the three models show distinct failure profiles on the same specifications
(Appendix~\ref{app:failuredist}, Table~\ref{tab:failures}). Configuration binding is the largest
category for all three, but its dominance varies. It accounts for 74\% of GPT-5's failures and
73\% of Opus 4.5's, while Gemini-2.5-pro splits nearly evenly between binding at 47\% and the
parser at 41\%, a share roughly triple that of the other two. At least two of the three models
produce the same outcome for 85 of the 100 specifications, and all three fail in the same category
on 30 (of which 29 are configuration-binding), so failure is driven as much by the specification as
by the model.

The failure mix also shifts with difficulty. Parse failures grow from 14\% of basic-tier outputs
to 24\% at the advanced tier, while configuration binding remains the largest category in every
tier (Table~\ref{tab:failure-by-difficulty}). Malformed-temporal failures peak at the intermediate tier
and nearly vanish at the advanced tier, where 91\% of outputs already fail at parsing or binding
and never reach semantic evaluation.

\subsection{Two Evaluation Modes}
\label{sec:twomodes}
Because every gold specification ships with its runnable configuration, \Ours supports two
evaluation modes on the same specifications. The default mode gives the model only the description,
testing whether it can both model the system and recover the exact interface. The
configuration-aware mode additionally supplies the constant and property names, isolating modeling
skill from interface matching. Table~\ref{tab:twomodes} reports both modes for all six models:
supplying the names raises correctness for every model and preserves the ranking, lifting the
strongest from 16\% to 26\% while the parse rate stays near 90\%, so much of the default-mode
failure is interface recovery rather than modeling. Even with the interface supplied the best model
tops out near 26\%, and the wide gap between a 90\% parse rate and a correct rate below 30\% is what
the exact oracle exposes and a parse-only benchmark would miss.

\subsection{Pass Quality}
\label{sec:passquality}
EQ3 asks whether the passes that models earn are meaningful. We apply two distinct probes to the
30 correct outputs pooled across the three frontier models, and because they test different
things they give different counts.

\emph{The substantive-pass screen} asks whether a pass exercises real behavior. A pass counts as
substantive when its specification reaches more than one state, the property it checks is not a
syntactic tautology, and that property is a behavioral one rather than a pure type or
well-formedness invariant. \textbf{12 of the 30} passes meet this bar. Of the other 18, none
check a literally tautological property. Nine never leave their initial state, unlike their
references, which the fixture filter of Section~\ref{sec:setup} guarantees are multi-state. The
remaining nine check only a type invariant, which holds by construction and rules out nothing
about the system's behavior. The plain correct rate therefore overstates genuine capability by
roughly 2.5 times.

\emph{The mutation test} asks whether the checked property is load-bearing. For each passing
output we replace every checked safety invariant named by the configuration with the literal
\texttt{TRUE} and re-run TLC. When the configuration checks a temporal property rather than a
mutable safety invariant, the test does not apply and the pass is scored as non-surviving, so the
test is a lower bound on vacuity. Of the 30 passes, 12 (distinct from the twelve substantive passes
above) check a temporal property with no mutable safety invariant and are scored non-surviving on
that ground, while 18 name a safety invariant the test can act on. A pass \emph{survives} when the mutated run still
fails, which shows the property genuinely constrains behavior. A pass whose mutated run then
succeeds was vacuous. \textbf{5 of the 18 testable passes} survive, an even stricter cut than the
substantive-pass screen because it demands the specific property do work, not merely that the module
move. Counted against all 300 outputs this is 5 of 300, the 1.7\% envelope bound; because the 12
temporal-only passes are scored against the rate rather than excluded, that 1.7\% is a conservative
lower bound rather than a direct vacuity measurement. We report the
mutation result where it bears on oracle scope in Section~\ref{sec:limitations}.

The two counts, 12 substantive and 5 mutation-surviving, are consistent rather than
contradictory. They measure different bars, and the stricter mutation test, which demands the
named property itself do work, admits fewer passes than the substantive screen, which only asks
that the module move and avoid a tautology. Reporting pass quality alongside the raw rate shows
how often a model produces a specification that both binds and verifies something real.
\section{Limitations}
\label{sec:limitations}

Several things this paper does \emph{not} establish. The descriptions are model-generated and
audited faithful (Section~\ref{sec:quality}), not a substitute for descriptions elicited from
practitioners requesting a specification from scratch. We draw one sample per specification, so the
rates carry binomial sampling error of roughly $\pm$4 to 7 points, small frontier differences are
indicative, and pass@k under an exact oracle is left to future work. The main results grade every
model on the GPT-5-written declarative descriptions, holding the input fixed but leaving the ranking
under the second provider untested; the dataset ships both so the check can be run, though the
providers had different length guidance, so the provider axis is confounded with prompt wording. The
name-hidden intent description ships for every specification as a control for identifier recall, but
the evaluation uses only the declarative style, so we make no contamination-controlled claim and
leave the intent run to future work (Section~\ref{sec:future}). The difficulty-label reliability
check is a 50-specification sample with non-blind annotators
(Appendix~\ref{sec:labelreliability}), defending the labels against arbitrariness rather than
establishing a direct measure of difficulty. Finally, composition and refinement, where modeling is
hardest, are under-represented in the evaluation set, which draws only 17 of its 100 from the 178
multi-module gold specifications.

Two scope limits are bounds we measure rather than only state, both regimes of the correctness
envelope (Section~\ref{sec:envelope}): the oracle is property-scoped, so a weaker-than-intended
property can pass, which the substantive and mutation regimes bound; and the dominant failure mode
is harness-coupled, since configuration-binding conflates modeling error with name mismatch, bounded
by the configuration-aware regime. A reference-varying behavioral gate is future work.
\vspace{-1.02em}
\section{Future Work}
\label{sec:future}

The dataset supports a broader set of studies, catalogued in full in
Appendix~\ref{sec:rq-catalog}; four are most immediate. A \emph{name-leakage} study over the paired
declarative and intent descriptions would measure how much correctness drops when identifier names
are hidden~\cite{whennames,varbench}, the study the shipped intent style is built for and the one
axis the present evaluation leaves unexercised. \emph{Reference-mutation grading} would extend the
pass-quality probes by perturbing the reference and requiring the checker to detect the change,
crediting correctness only to specifications that constrain behavior against an independently varied
ground truth~\cite{beer2001vacuity,verina}, lifting the restriction that makes our mutation test a
lower bound. A \emph{contamination-over-time} study would
use the recorded repositories and a recency split to test whether accuracy tracks repository
age~\cite{livecodebench,contamsurvey}. A \emph{cross-provider} study would hold the specification
fixed and vary the description's author, isolating the prompt-source effect from the solving
model~\cite{livebench}.
\section{Conclusion}
\label{sec:conclusion}

\Ours grades natural-language to TLA\textsuperscript{+} generation by execution rather than
resemblance, over 403 model-checked gold and 897 parse-only silver specifications from 13
repositories. Its central lesson is that an exact oracle does not settle correctness on its own: the
same outputs admit a range of defensible rates, from 18.7\% down to 1.7\%, and naming that
correctness envelope is what lets a benchmark report where a model sits rather than a single silent
number. Inside the envelope the findings are stable, with every model far better at valid than
correct TLA\textsuperscript{+} and a sharp fall in correctness with difficulty. We release the
specifications, descriptions, outputs, and grading tools so these measurements can be reproduced and
extended.

\clearpage
\bibliographystyle{ACM-Reference-Format}
\bibliography{references}

\clearpage
\appendix
\renewcommand{\thesection}{A\arabic{section}}
\renewcommand{\thetable}{A\arabic{table}}
\renewcommand{\thefigure}{A\arabic{figure}}
\setcounter{table}{0}
\setcounter{figure}{0}

\section*{Appendix}
\addcontentsline{toc}{section}{Appendix}

\section{Ethics}
\label{sec:ethics}

The specifications are third-party material from 13 public repositories, and each specification
records its source repository, its URL, and its path, so the original license applies and can be
traced. All 13 repositories carry a license that permits redistribution with attribution, so every
specification is re-hosted unchanged with its recorded checksum; we include each repository's
license in the released manifest. Our own contributions, namely the descriptions, the labels, and the manifest, are
released under CC BY 4.0.

The two annotation studies in this paper, the faithfulness audit of
Section~\ref{sec:quality} and the difficulty relabeling of Section~\ref{sec:difficulty}, were
carried out by members of the author team and close collaborators rather than by recruited
participants. We collected no personal data about them and we release no annotator identifiers,
so no ethics-board review applies. We disclose the arrangement in both places because it means
neither study is blind to authorship.

We ask that the dataset not be used as training data. It is an evaluation resource, and training
on it would defeat its purpose and contaminate future evaluations that rely on it. A permissive
license cannot enforce this request, so the release embeds a canary string that makes later
contamination of the benchmark detectable.

\paragraph{Use of large language models.} Large language models are the object of study here, and
their role in producing the dataset, namely writing the natural-language descriptions, is described
in Section~\ref{sec:construction}. Separately, in preparing the manuscript itself the authors used
a large language model only to improve wording and to assist with figures and code. Every
suggestion was reviewed by the authors, who take full responsibility for the findings and all
content in this paper.
\section{Reproducibility and Accessibility}
\label{sec:repro}

We release the full dataset, namely the gold and silver specifications with their runnable
configurations, the four description sets, and a manifest recording every specification's tier,
labels, provenance, and file checksum. The prompts are reproduced in
Appendix~\ref{sec:prompts} and a lifecycle datasheet in Appendix~\ref{sec:datasheet}. Grading is
defined in Section~\ref{sec:grading} over the standard TLA\textsuperscript{+} tools, so a user
with those tools reproduces the verdict for any generated specification.

The release also carries what each bound of the correctness envelope needs. The fixture flags in
the manifest set the pruning choice, the runnable configurations support the
configuration-aware mode, and the gold configurations that name real properties give the
mutation test something to act on. We release the model outputs alongside the specifications, so
the default, substantive, and mutation rows of Table~\ref{tab:envelope} can be recomputed from the
released outputs without querying a model; the configuration-aware row is a separate set of
generations produced by re-querying with the interface names supplied, and we release those
outputs as well.

The model checker is run with a single worker at the Java default heap under a 300-second
per-specification wall-clock limit, with SANY parsing bounded at 60 seconds, and state explosion is
bounded by that time limit rather than an explicit state cap. A user running the released grader
reproduces every table on comparable hardware, with the resource-limit category dependent on the
stated time and memory budget, since a run that exhausts the 300-second limit on slower hardware or
a smaller heap is recorded as a resource limit. Deadlock checking is left enabled, its TLC default,
and each gold reference is verified under this budget as part of tier assignment
(Section~\ref{sec:construction}), so a deadlock outcome on a generated output reflects a model error
rather than a harness default.

Each model was queried once per specification with a fixed generation budget of 16{,}000 tokens.
The three frontier models were the hosted \texttt{gpt-5}, \texttt{claude-opus-4-5}, and
\texttt{gemini-2.5-pro} endpoints, and the open models were \texttt{qwen2.5-coder-32b},
\texttt{llama-3.3-70b}, and \texttt{gpt-oss-20b} served locally through Ollama. The frontier models
were queried in July 2026 and the open models in late June 2026. The hosted endpoints were not
pinned to dated snapshots, so a later query may return a different underlying model; we release the
generated outputs so the graded results do not depend on re-querying. Decoding settings follow each
provider's interface. Gemini-2.5-pro was decoded greedily at
temperature 0. GPT-5 was decoded at its default temperature, since the reasoning model rejects
an explicit temperature of 0. Claude Opus 4.5 and the open models were decoded at their provider
default temperature. Because we draw a single sample per specification, the reported rates carry
binomial sampling error, quantified in Section~\ref{sec:limitations}; reporting the full sampling
distribution over multiple samples per specification is left to future work.

The release is archived on Zenodo under the persistent identifier
\texttt{10.5281/zenodo.21310317} so that a specific version can be
cited~\cite{tlabench_zenodo}. The grading tools and the code used to produce the results in this
paper are available at \url{https://github.com/LUC-AI4FM/tla_benchmark/tree/reviewer-release}.
\section{A Benchmark Instance}
\label{app:instance}

Figure~\ref{fig:instance} shows one complete instance from \Ours, referenced from
Section~\ref{sec:format}. It gives the gold specification with its configuration, and one
description of each style for the same specification.

\begin{figure}[h]
\centering
\begin{promptboxf}[{Gold specification and configuration}]{goldcolor}
\ttfamily\scriptsize
\begin{verbatim}
----------- MODULE HourClock -----------
EXTENDS Naturals
VARIABLE hr
HCini == hr \in (1 .. 12)
HCnxt == hr' = IF hr # 12 THEN hr + 1 ELSE 1
HC    == HCini /\ [][HCnxt]_hr
----------------------------------------
THEOREM HC => []HCini
========================================
\end{verbatim}
cfg:\ \ SPECIFICATION HC\ \ \ \ INVARIANT HCini
\end{promptboxf}
\vspace{4pt}
\begin{promptboxf}[{Declarative description (name-revealing)}]{declcolor}
\scriptsize
The module models a simple 12-hour clock. The single variable \texttt{hr} represents the
current hour. The initial condition \texttt{HCini} constrains \texttt{hr} to be a natural
number in the range 1 through 12. The next-state relation \texttt{HCnxt} advances the clock by
one hour each step and wraps from 12 back to 1. The configuration checks \texttt{HCini} as an
invariant.
\end{promptboxf}
\vspace{4pt}
\begin{promptboxf}[{Intent description (name-hidden)}]{intentcolor}
\scriptsize
Write a TLA+ specification for a simple 12-hour clock. The clock maintains a single hour
counter that always holds a value between 1 and 12 inclusive. Each tick advances the hour by
one and wraps from 12 back to 1. The system begins in any valid hour state and transitions by
this cyclic increment rule, with stuttering steps permitted. The key correctness property is
that the hour value remains within 1 through 12 at all times.
\end{promptboxf}
\caption{One benchmark instance. The gold specification and its configuration (top), and one
description of each style for the same specification. The declarative style keeps the module's
names and the intent style hides them. Labels for this specification are difficulty
\emph{basic} and category \emph{system}. This is the same specification used in the worked
failure of Figure~\ref{fig:worked} and Appendix~\ref{sec:error-examples}, so a reader can follow one
instance from its description through to a graded model output.}
\label{fig:instance}
\end{figure}
\section{Provenance and Exclusions}
\label{sec:provenance-appendix}

\begin{figure*}[t]
\centering
\includegraphics[width=\textwidth]{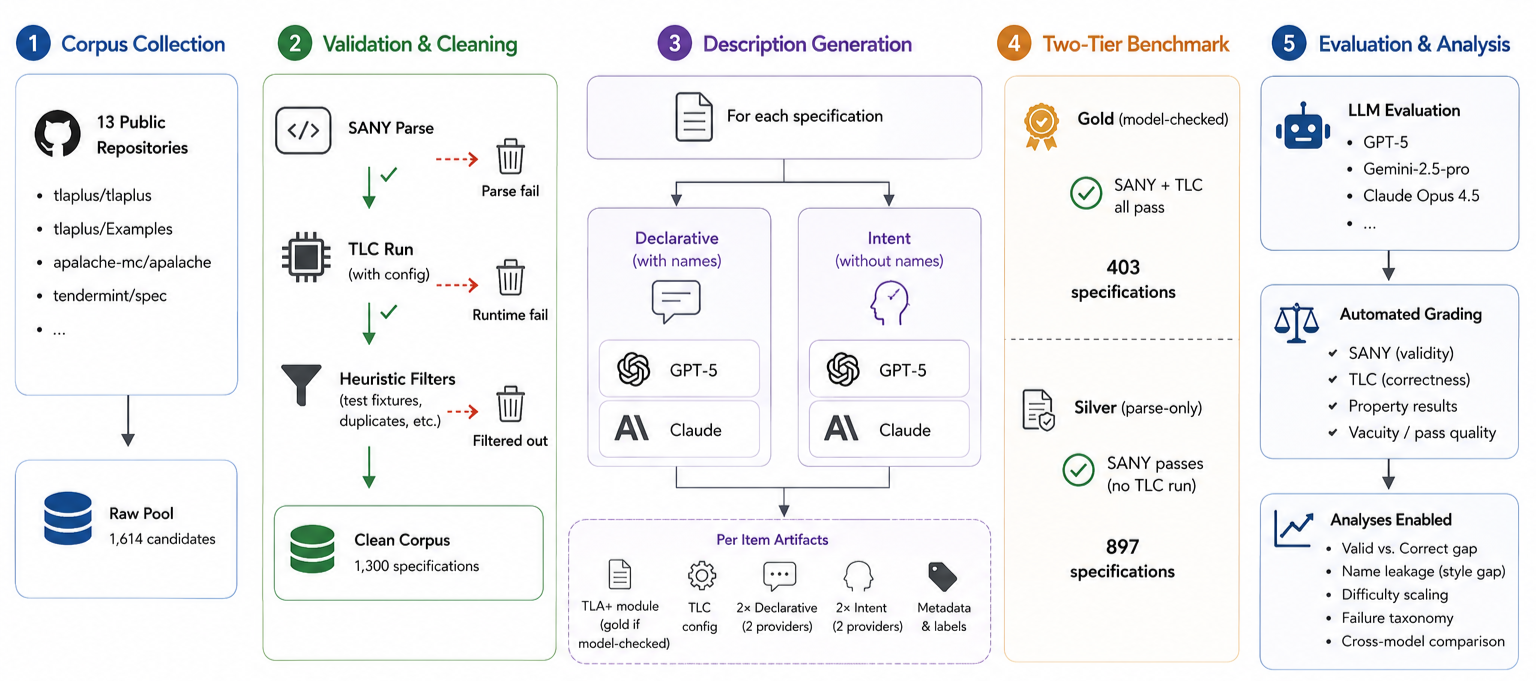}
\caption{Construction and evaluation pipeline. We collect specifications from 13 public
repositories (1{,}614 candidates), clean them with SANY parsing, TLC model checking, and
heuristic filters (1{,}300 specifications), generate two description styles from two providers
for each specification, split the corpus into a model-checked gold tier and a parse-only silver
tier, and evaluate models with automated SANY and TLC grading.}
\label{fig:pipeline}
\end{figure*}

Table~\ref{tab:provenance} lists the specifications contributed by each source repository,
split by tier and by difficulty. The dataset draws from 13 public repositories across 9 GitHub
organizations. Two repositories, \texttt{apalache-mc/apalache} and \texttt{tlaplus/tlaplus},
contribute 776 of the 1{,}300 specifications between them, so the corpus is concentrated in the
largest public TLA\textsuperscript{+} collections even though it spans 13 sources.

\begin{table}[h]
\centering
\caption{Specifications per source repository, by tier and difficulty, sorted by total
contribution. The 13 repositories span 9 GitHub organizations.}
\label{tab:provenance}
\footnotesize
\setlength{\tabcolsep}{3.5pt}
\renewcommand{\arraystretch}{1.12}
\begin{tabular}{@{}lrrrrrr@{}}
\toprule
\textbf{Source repository} & \textbf{Gold} & \textbf{Silver} & \textbf{Total} &
\textbf{Basic} & \textbf{Interm.} & \textbf{Adv.} \\
\midrule
\texttt{apalache-mc/apalache}      & 84 & 341 & 425 & 327 & 64 & 34 \\
\texttt{tlaplus/tlaplus}           & 140 & 211 & 351 & 197 & 89 & 65 \\
\texttt{tlaplus/Examples}          & 123 & 176 & 299 & 122 & 84 & 93 \\
\texttt{josehu07/learn-tla}        & 35 & 31 & 66 & 28 & 25 & 13 \\
\texttt{tlaplus/CommunityModules}  & 5 & 57 & 62 & 42 & 13 & 7 \\
\texttt{tendermint/spec}           & 0 & 49 & 49 & 32 & 6 & 11 \\
\texttt{pingcap/tla-plus}          & 9 & 9 & 18 & 8 & 1 & 9 \\
\texttt{microsoft/CCF}             & 6 & 10 & 16 & 8 & 4 & 4 \\
\texttt{atomix/atomix-tlaplus}     & 0 & 6 & 6 & 2 & 1 & 3 \\
\texttt{Azure/azure-cosmos-tla}    & 0 & 3 & 3 & 0 & 0 & 3 \\
\texttt{TeamTilapia/vscode-tilapia}& 1 & 1 & 2 & 0 & 0 & 2 \\
\texttt{tlaplus/ValidationTestSuite}& 0 & 2 & 2 & 0 & 2 & 0 \\
\texttt{tlaplus/DrTLAPlus}         & 0 & 1 & 1 & 0 & 0 & 1 \\
\midrule
\textbf{Total} & \textbf{403} & \textbf{897} & \textbf{1{,}300} &
\textbf{766} & \textbf{289} & \textbf{245} \\
\bottomrule
\end{tabular}
\end{table}

\paragraph{Excluded specifications.}
The crawl, after removal of exact duplicates, yielded 1{,}614 candidate modules, and the
cleaning pipeline removed 314 of them before the 1{,}300 release (Table~\ref{tab:excluded}).
The pipeline ran over the pooled candidate set rather than repository by repository, so the drop
counts are corpus-wide and cannot be attributed to a source. The per-repository counts in
Table~\ref{tab:provenance} are therefore counts of released specifications, and the manifest
records the source of every one of them.

\begin{table}[h]
\centering
\caption{Specifications excluded during cleaning (corpus-wide), largest category first.}
\label{tab:excluded}
\small
\renewcommand{\arraystretch}{1.15}
\begin{tabular}{@{}lr@{}}
\toprule
\textbf{Exclusion reason} & \textbf{Count} \\
\midrule
Near-duplicate / non-conforming  & 101 \\
SANY parse failure               & 88 \\
Tooling / counterexample modules & 68 \\
Test scaffolding                 & 42 \\
Runtime failure                  & 15 \\
\midrule
\textbf{Total excluded}          & \textbf{314} \\
\midrule
Raw candidates                   & 1{,}614 \\
Released (gold + silver)         & 1{,}300 \\
\bottomrule
\end{tabular}
\end{table}
\section{Additional Tables}
\label{app:failuredist}

This appendix collects tables referenced from the main text. They are the corpus breakdown by
difficulty tier (Table~\ref{tab:difficulty}, from Section~\ref{sec:stats}), the full per-model
failure distribution over the taxonomy (Table~\ref{tab:failures}, from
Section~\ref{sec:fingerprints}), the failure-by-difficulty cross-tabulation
(Table~\ref{tab:failure-by-difficulty}, from Section~\ref{sec:fingerprints}), the
difficulty-stratified parse and correct rates (Table~\ref{tab:difficulty-results}, from
Section~\ref{sec:difficulty}), and the default-versus-configuration-aware comparison
(Table~\ref{tab:twomodes}, from Section~\ref{sec:twomodes}). The tables appear in this order.

\begin{table}[h]
\centering
\caption{\Ours by difficulty tier. Size is non-comment lines of code. Mean size grows steeply
across tiers, so the labels correlate strongly with specification size.
Section~\ref{sec:difficulty} reports the inter-annotator reliability check and the
construct-derived robustness check that defend the labels beyond this gradient.}
\label{tab:difficulty}
\small
\renewcommand{\arraystretch}{1.15}
\begin{tabular}{@{}lrrrrr@{}}
\toprule
\textbf{Difficulty} & \textbf{Specs} & \textbf{Gold} & \textbf{Silver} &
\makecell{\textbf{LOC}\\\textbf{mean}} & \makecell{\textbf{LOC}\\\textbf{median}} \\
\midrule
Basic        & 766   & 260 & 506 & 17.1  & 13  \\
Intermediate & 289   & 83  & 206 & 58.0  & 48  \\
Advanced     & 245   & 60  & 185 & 208.8 & 109 \\
\midrule
\textbf{Total} & \textbf{1{,}300} & \textbf{403} & \textbf{897} & \textbf{62.3} & \textbf{23} \\
\bottomrule
\end{tabular}
\end{table}

\begin{table}[t]
\centering
\caption{Failure distribution over the taxonomy on the 100-specification evaluation set, by model. Each
column counts outputs of that model. Models fail in visibly different ways on the same specifications.}
\label{tab:failures}
\small
\setlength{\tabcolsep}{5pt}
\renewcommand{\arraystretch}{1.12}
\begin{tabular}{@{}lrrr@{}}
\toprule
\textbf{Outcome} & \textbf{GPT-5} & \textbf{Gemini-2.5-pro} & \textbf{Opus 4.5} \\
\midrule
Correct                 &  4 & 10 & 16 \\
\midrule
Config-binding failure  & 71 & 42 & 61 \\
Parse failure           & 11 & 37 & 13 \\
Malformed temporal      & 12 &  6 &  3 \\
Deadlock                &  1 &  2 &  3 \\
Safety violation        &  0 &  1 &  2 \\
Liveness violation      &  0 &  0 &  0 \\
Run-time error          &  0 &  1 &  1 \\
Resource limit          &  1 &  1 &  1 \\
\midrule
\textbf{Total}          & \textbf{100} & \textbf{100} & \textbf{100} \\
\bottomrule
\end{tabular}
\end{table}
\begin{table}[t]
\centering
\caption{Outcome category by difficulty tier, pooled across the three frontier models on the
100-specification evaluation set, for 300 outputs. Column sums are the per-tier output counts,
which are three models times 35, 34, and 31 specifications. Configuration binding stays the
dominant failure in every tier. Counts use the frozen grading of Section~\ref{sec:grading},
under which a missing configuration counts as a configuration-binding failure, so the
configuration-binding row reconciles with Table~\ref{tab:failures}.}
\label{tab:failure-by-difficulty}
\small
\renewcommand{\arraystretch}{1.12}
\begin{tabular}{@{}lrrr@{}}
\toprule
\textbf{Outcome} & \textbf{Basic} & \textbf{Inter.} & \textbf{Adv.} \\
\midrule
Correct                & 26 &  2 &  2 \\
\midrule
Parse failure          & 15 & 24 & 22 \\
Config-binding failure & 53 & 58 & 63 \\
Malformed temporal     &  6 & 14 &  1 \\
Safety violation       &  1 &  0 &  2 \\
Run-time error         &  0 &  2 &  0 \\
Deadlock               &  2 &  2 &  2 \\
Resource limit         &  2 &  0 &  1 \\
\midrule
\textbf{Total}         & \textbf{105} & \textbf{102} & \textbf{93} \\
\bottomrule
\end{tabular}
\end{table}

\begin{table}[t]
\centering
\caption{Parse (P) and correct (C) rate within each difficulty stratum on the 100-specification
evaluation set with the declarative description, under the default regime. Denominators are 35,
34, and 31 per stratum, and 105, 102, and 93 for the pooled cells. The pooled row matches the
cliff quoted in Section~\ref{sec:difficulty}.}
\label{tab:difficulty-results}
\small
\setlength{\tabcolsep}{5pt}
\renewcommand{\arraystretch}{1.12}
\begin{tabular}{@{}lrrrrrr@{}}
\toprule
& \multicolumn{2}{c}{\textbf{Basic}} & \multicolumn{2}{c}{\textbf{Inter.}} & \multicolumn{2}{c}{\textbf{Adv.}} \\
\cmidrule(lr){2-3}\cmidrule(lr){4-5}\cmidrule(lr){6-7}
\textbf{Model} & \textbf{P} & \textbf{C} & \textbf{P} & \textbf{C} & \textbf{P} & \textbf{C} \\
\midrule
Claude Opus 4.5 & 83\% & 40\% & 94\% & 3\% & 84\% & 3\% \\
Gemini-2.5-pro  & 83\% & 26\% & 47\% & 3\% & 58\% & 0\% \\
GPT-5           & 91\% &  9\% & 88\% & 0\% & 87\% & 3\% \\
\midrule
Pooled          & 86\% & 25\% & 76\% & 2\% & 76\% & 2\% \\
\bottomrule
\end{tabular}
\end{table}

\begin{table}[t]
\centering
\caption{Default versus configuration-aware correct rate per model. The two modes are defined in
Section~\ref{sec:twomodes}, where the configuration-aware mode additionally supplies the
constant and property names the configuration uses. Default rates are the correct column of
Table~\ref{tab:main}. Supplying the interface names raises correctness for every model while the
ranking is preserved, and the best model still tops out at 26\%. Pooled over the three frontier
models these columns give the default and configuration-aware rows of
Table~\ref{tab:envelope}.}
\label{tab:twomodes}
\small
\renewcommand{\arraystretch}{1.12}
\begin{tabular}{@{}lrr@{}}
\toprule
\textbf{Model} & \textbf{Default} & \textbf{Config-aware} \\
\midrule
\multicolumn{3}{@{}l}{\textit{Frontier}}\\
Claude Opus 4.5    & 16\% & 26\% \\
Gemini-2.5-pro     & 10\% & 19\% \\
GPT-5              &  4\% & 11\% \\
\midrule
\multicolumn{3}{@{}l}{\textit{Open}}\\
qwen2.5-coder-32b  &  1\% &  4\% \\
llama3.3-70b       &  0\% &  2\% \\
gpt-oss-20b        &  1\% &  3\% \\
\bottomrule
\end{tabular}
\end{table}
\section{A Worked Failure in Full}
\label{sec:error-examples}

To make the grading concrete, we reproduce the evaluation case of Figure~\ref{fig:worked} in
full, adding what the main text omits, namely the complete model-checker log, verbatim. The
offending line in the model output and the error line in the log are marked in
\textcolor{nored}{\textbf{bold red}}. This is the same specification shown as a complete
benchmark instance in Appendix~\ref{app:instance}, so the description, the reference, and a
graded model output can be read together.

\needspace{10\baselineskip}
\paragraph{Reference specification (gold).}
\begin{lstlisting}[style=tla]
----------- MODULE HourClock -----------
EXTENDS Naturals
VARIABLE hr
HCini == hr \in (1 .. 12)
HCnxt == hr' = IF hr # 12 THEN hr + 1 ELSE 1
HC    == HCini /\ [][HCnxt]_hr
----------------------------------------
THEOREM HC => []HCini
========================================
\end{lstlisting}
The configuration checks \texttt{HCini} as an invariant of \texttt{HC} and declares no
constants.

\needspace{9\baselineskip}
\paragraph{Model output (GPT-5).}
\begin{lstlisting}[style=tla]
----------- MODULE HourClock -----------
EXTENDS Integers
@@CONSTANTS Hrs@@
@@ASSUME Hrs = 1..12@@
VARIABLES hr
HCini == hr \in Hrs
HCnxt == hr' = IF hr # 12 THEN hr + 1 ELSE 1
Spec  == HCini /\ [][HCnxt]_hr
========================================
\end{lstlisting}
The output parses under SANY, so it is valid TLA\textsuperscript{+}. It differs from the
reference in two binding-relevant ways. First, the model introduces a declared constant
\texttt{Hrs} and an assumption about it, rather than writing the range \texttt{1 .. 12} inline as
the reference does; because the reference configuration declares no constants, it assigns no value
to \texttt{Hrs}, so the model checker cannot instantiate the module. Second, the model names its
top-level formula \texttt{Spec} where the configuration expects \texttt{HC}, a further binding
mismatch. Either one alone blocks binding under the default regime.

\needspace{9\baselineskip}
\paragraph{Model-checker log (TLC), verbatim.}
\begin{lstlisting}[style=tla]
TLC2 Version 2026.04.22
Running breadth-first search Model-Checking ...
Semantic processing of module Naturals
Semantic processing of module Integers
Semantic processing of module HourClock
Linting of module HourClock
Starting...
@@Error: The constant parameter Hrs is not
  assigned a value by the configuration file.@@
Finished in 00s
\end{lstlisting}
This is a configuration-binding failure. The model produced a well-formed module whose only
defect, relative to the fixed configuration, is the names and parameters it chose. The failure
is a mismatch between the module's interface and the interface the configuration expects, not a
failure to parse. The module's clock transition logic matches the reference's, differing only in
the interface names, which is why we classify this failure as binding rather than modeling. This case is also the clearest
illustration of why the correctness envelope has an interface bound. Under the default regime
this output is simply wrong, while under the configuration-aware mode of
Section~\ref{sec:twomodes} the model is given the names the configuration expects and the same clock
logic is graded on its merits.
\section{Catalog of Research Questions Enabled by the Dataset}
\label{sec:rq-catalog}

Table~\ref{tab:rqcatalog} lists research questions the dataset lets others answer. Each is tied
to a specific finding or an explicitly stated future-work or limitations direction in prior
work, and the \emph{Source} column gives that citation. Questions answerable from released data
are tagged \textsc{[c]} (computed), and those needing new inference or checking are tagged
\textsc{[r]} (needs-runs). The \emph{Asset} column names the part of the dataset that enables
the question.

This paper answers four of them, which shows the catalog is a description of what the resource
supports rather than a list of intentions. RQ10 is Section~\ref{sec:validity}, RQ14 is the
pass-quality analysis of Section~\ref{sec:passquality}, RQ17 is the correctness cliff of
Section~\ref{sec:difficulty}, and RQ20 is the failure fingerprints of
Section~\ref{sec:fingerprints}. RQ5 is answered in part, since
Section~\ref{sec:construction} reports that 46 of the 146 system-category gold specifications
are trivial fixtures, while the effect of pruning on measured accuracy is left for a user to
compute from the released flags. The remaining questions need work this paper does not do.

\begin{table*}[t]
\centering
\caption{Research questions enabled by the dataset, grouped by theme. \textsc{[c]}~=~computable
from released data, \textsc{[r]}~=~needs new runs. Every row is backed by a cited source. The
five rows marked \dag\ are answered in whole or in part by this paper.}
\label{tab:rqcatalog}
\small
\renewcommand{\arraystretch}{1.25}
\begin{tabular}{@{}p{0.5cm} p{7.6cm} p{3.1cm} p{2.0cm} c@{}}
\toprule
\textbf{\#} & \textbf{Research question} & \textbf{Dataset asset} & \textbf{Source} & \textbf{Tag} \\
\midrule
\multicolumn{5}{@{}l}{\textit{\textbf{Contamination}}}\\
RQ1 & Does accuracy decline with source-repository release date? & Dated repos + exact oracle & \cite{livecodebench,contamsurvey} & \textsc{[r]} \\
RQ2 & Can an exact model-checker signal be a contamination-robust metric where paraphrase decontamination fails? & Oracle + paraphrase descriptions & \cite{contamsurvey} & \textsc{[r]} \\
RQ3 & Does the static-to-dynamic overfitting effect reproduce in \TLA? & $2{\times}2$ description matrix & \cite{contamstatic} & \textsc{[r]} \\
RQ4 & Do CDD/TED contamination detectors extend to formal-spec generation? & Oracle + multi-sample & \cite{genmemorize} & \textsc{[r]} \\
RQ5\dag & What fraction of a public corpus is tooling/test-fixture material, and how much does pruning move accuracy? & Fixture-flagged sources & \cite{mbpp,openclassgen} & \textsc{[c]}/\textsc{[r]} \\
\midrule
\multicolumn{5}{@{}l}{\textit{\textbf{Memorization and Name Leakage}}}\\
RQ6 & Does hiding names (intent vs.\ declarative) reduce correctness though the task is unchanged? & Name-leakage axis & \cite{whennames} & \textsc{[r]} \\
RQ7 & Does the GSM8K-style perturbation drop appear in \TLA? & Intent as perturbation & \cite{varbench} & \textsc{[r]} \\
RQ8 & Do models succeed by memorizing lexically-similar specs (novel-structure split)? & Spec text + source provenance & \cite{leandojo} & \textsc{[r]} \\
RQ9 & Does pre/post-obfuscation reporting reorder model rankings? & Declarative/intent pair & \cite{whennames,livebench} & \textsc{[r]} \\
\midrule
\multicolumn{5}{@{}l}{\textit{\textbf{Oracle: Validity vs.\ Correctness}}}\\
RQ10\dag & How large is the parse-vs-correct gap, and does ``checker passed'' overstate correctness? & SANY+TLC per spec & \cite{sysmobench,verina} & \textsc{[c]} \\
RQ11 & Can TLC counterexample traces seed a semantic-error detection benchmark? & Traces + failure taxonomy & \cite{nl2sqlbugs} & \textsc{[c]}/\textsc{[r]} \\
RQ12 & Does exact correctness diverge from surface similarity (BLEU/embedding)? & Exact labels + text & \cite{proofnet,openclassgen} & \textsc{[c]} \\
RQ13 & Does SANY parse rate predict TLC correctness (cheap predictor)? & Parse+correct outcomes & \cite{proofnet} & \textsc{[c]} \\
\midrule
\multicolumn{5}{@{}l}{\textit{\textbf{Metric Design and Pass Quality}}}\\
RQ14\dag & How prevalent are tautological/degenerate specs that pass but assert nothing? & Pass-quality axis & \cite{beer2001vacuity,kupferman2003vacuity} & \textsc{[c]} \\
RQ15 & Does reference-perturbing mutation scoring separate meaningful specs from trivial passes? & Gold specs + oracle & \cite{beer2001vacuity,verina} & \textsc{[r]} \\
RQ16 & How does pass@k scale under an exact formal oracle? & Oracle + multi-sample & \cite{humaneval} & \textsc{[r]} \\
\midrule
\multicolumn{5}{@{}l}{\textit{\textbf{Difficulty and Scaling}}}\\
RQ17\dag & Is there a correctness cliff by difficulty, and does difficulty (not date) drive it? & Difficulty labels & \cite{swebench,dafnybench} & \textsc{[c]} \\
RQ18 & Does accuracy scale log-linearly with model size, or is size non-predictive? & Oracle across sizes & \cite{mbpp,formallm} & \textsc{[r]} \\
RQ19 & Do code-specialized models underperform via negative transfer? & Oracle across families & \cite{formallm} & \textsc{[r]} \\
\midrule
\multicolumn{5}{@{}l}{\textit{\textbf{Failure Analysis}}}\\
RQ20\dag & Do models show distinct failure fingerprints on the same specifications? & 8-class taxonomy & \cite{formallm} & \textsc{[c]} \\
RQ21 & Are models weaker at temporal/liveness than safety properties? & Safety/liveness classes & \cite{sysmobench} & \textsc{[c]} \\
RQ22 & Does correctness degrade as the number of constants, variables, and operators grows? & Spec text + oracle outcomes & \cite{humaneval} & \textsc{[c]} \\
\midrule
\multicolumn{5}{@{}l}{\textit{\textbf{Cross-Provider and Description Quality}}}\\
RQ23 & Does the description provider (GPT-5 vs.\ Claude) change downstream correctness? & Two providers per spec & \cite{livebench,autoformsurvey} & \textsc{[r]} \\
RQ24 & Is the dominant failure a missing load-bearing assumption (intent-to-formal misalignment)? & Intent + traces & \cite{autoformllm} & \textsc{[r]} \\
\midrule
\multicolumn{5}{@{}l}{\textit{\textbf{Repair and Documentation}}}\\
RQ25 & Can models repair a failing spec from the TLC counterexample trace? & Traces as feedback & \cite{dafnybench} & \textsc{[r]} \\
RQ26 & Can others recreate an equivalent contamination-free benchmark from our pipeline? & Documented pipeline & \cite{datasheets} & \textsc{[c]} \\
\bottomrule
\end{tabular}
\end{table*}
\section{Relevance to the Data Science and Knowledge-Discovery Community}
\label{sec:relevance}

\Ours is a datasets-and-benchmarks contribution, and we outline here why it is of interest to
the broader KDD community rather than only to the formal-methods audience.

\textbf{An execution-grounded benchmark for a generation task.}
The central methodological artifact is an exact, automatic correctness oracle for a
natural-language-to-structured-artifact generation task. This is the same evaluation problem
that recurs across data science whenever a model produces an executable artifact from a
description, for example natural-language to SQL, to code, or to query plans, where the field
has repeatedly found that approximate oracles based on textual overlap or a finite test set
overstate capability. \Ours provides a setting where the oracle is decidable rather than
approximate. The construction recipe also transfers. Any domain that ships an executable ground
truth with checkable properties can pair each instance with a fixed reference configuration and
grade generated artifacts by running the checker over the full state space against the
properties that configuration names. The NL2SQL semantic-error line~\cite{nl2sqlbugs} that has
appeared at this venue is the closest analogue, and \Ours extends that concern from query
translation to full specification generation.

\textbf{The correctness envelope transfers further than the oracle does.}
The oracle is specific to TLA\textsuperscript{+}, but the finding it produces is not. Any
benchmark that grades a generated artifact by running it faces the same three choices we make
explicit in Section~\ref{sec:envelope}. It must decide which instances are worth grading, since
a corpus scraped from public sources carries degenerate cases that every model passes. It must
decide how much of the target interface the model is required to recover, since an artifact can
be functionally right and fail on a name. And it must decide whether a passing run is required
to check anything, since a test suite or a property set can be satisfied vacuously. A benchmark
that leaves these implicit reports one number from a range, and our range is a factor of eleven
wide, and sixfold on fixed outputs. We suggest that resource builders in adjacent settings report the range
rather than a point, and we release the assets that let a reader recompute ours.

\textbf{A resource with reusable structure for data-centric study.}
Beyond the single evaluation we report, the release is built to support data-centric
machine-learning research. It ships two description styles and two providers per specification,
difficulty labels and, for the gold tier, category labels, a model-checked gold tier and a
parse-only silver tier, per-file checksums, and a catalog of research questions
(Appendix~\ref{sec:rq-catalog}). These turn the input, the difficulty, the description provider,
and the grading strength into controllable variables, so a data-mining researcher can study
robustness, contamination, and data quality on a fixed corpus. The name-hidden intent style, in
particular, is shipped as a control that a user can apply to study how much of a model's
measured performance depends on identifier names rather than on the description, a question of
direct interest to anyone building or evaluating models on public-code corpora.

\textbf{Public, documented, and reproducible.}
The dataset is released openly on a persistent archive with a datasheet
(Appendix~\ref{sec:datasheet}), full provenance (Appendix~\ref{sec:provenance-appendix}), and
the open grading tools, so the resource meets the accessibility, documentation, and
reproducibility criteria the track asks for and can be picked up without contacting the authors.
\section{The Descriptions Are Not the Bottleneck}
\label{sec:notbottleneck}

A natural concern is that models fail because the descriptions are inadequate rather than because
the task is hard. We draw on the graded outputs of Section~\ref{sec:experiments}, and three
observations argue against the concern. First, the descriptions carry enough information to solve
the task: at least one model produces a fully correct specification for 19 of the 100 evaluation
specifications, and every specification is parsed by at least one model, so a workable description
exists even where other models fail on the same input. Second, the dominant failure is not
comprehension. Across the three frontier models, 64\% of the 270 failures are configuration-binding
failures, where the model understood the description well enough to build a well-formed module and
named its constants or properties differently from the fixed configuration; only 23\% are outright
parse failures, and a model that produces a well-formed but mis-named module has clearly read the
description. Third, the set of specifications each model gets right differs from model to model on the same
descriptions, and no single model solves all 19 that some model solves, which could not happen if
the description alone determined the result.
Together these observations place the difficulty in the models and the task rather than in the
descriptions.

\section{Difficulty-Label Reliability}
\label{sec:labelreliability}

To check the difficulty labels that drive the correctness cliff of Section~\ref{sec:difficulty},
two TLA\textsuperscript{+}-literate annotators independently relabeled a random 50-specification
sample under the same three-tier rubric. Their labels reach a Cohen's $\kappa$ of 0.8, substantial
agreement for a three-way judgment; because the tiers are ordinal, an ordinal-weighted agreement,
which penalizes an adjacent basic-versus-intermediate disagreement less than a
basic-versus-advanced one, rises to 0.85. The annotators are close collaborators rather than
independent third parties, so the check is not blind, a caveat we disclose as we do for the
faithfulness audit. As a robustness check we repeat the difficulty analysis with a measure derived
mechanically from the TLA\textsuperscript{+} constructs each specification uses, and the same
collapse appears, from 22\% to 4\%. Because the rubric is itself keyed to construct depth
(Section~\ref{sec:construction}), this measure shares a construct family with the labels, so it
tests whether the labels were applied consistently rather than supplying a fully independent notion
of difficulty; the cliff survives under both the hand labels and the mechanical proxy.

\section{Exact Prompts Given to the Language Models}
\label{sec:prompts}

This appendix reproduces the prompts behind the dataset and the evaluation. The two description
prompts below are the versions issued to Claude Opus 4.5, reproduced verbatim from the release
scripts. The GPT-5 descriptions were generated from the same templates but with shorter length
guidance, which is why the GPT-authored descriptions are shorter on average than the
Claude-authored ones in Table~\ref{tab:stats}. Because the wording differs by provider, the
provider axis is confounded with the prompt, a limitation we state in
Section~\ref{sec:limitations}.

The exact GPT-5 prompts were not recorded in the release scripts. We reconstruct them as the
templates below with a length guidance of \textbf{80 to 120} words rather than the 120 to 220
issued to Claude. We flag this reconstruction explicitly because the GPT-5 declarative
descriptions are the evaluation input of Section~\ref{sec:experiments}, so a user reproducing
the corpus from these templates may obtain descriptions that differ in length from the released
ones. The released descriptions themselves are fixed and archived, so the evaluation in this
paper is reproducible from the release even though the prompt that produced them is not
recorded exactly.

The generation prompt is shared across all evaluated models and is reproduced verbatim. The
token ``\{description\}'' marks where the natural-language description is substituted.

\paragraph{Declarative description prompt (Claude version, verbatim).}
\begin{promptbox}[Declarative Description]{declcolor}
\ttfamily\small
You are a TLA+ specification expert. You will be given the source of a TLA+ module (and its
model-checking config, if any). Write a faithful, declarative description of what the module
specifies. Cover, when present, the system or algorithm being modeled, the CONSTANTS and
VARIABLES, the key actions or next-state relation, the temporal specification and any fairness
conditions, and the safety and liveness properties or invariants checked. Describe what the
spec is and does. Do not editorialize or suggest changes. Be faithful: do not claim properties
the module does not contain, and do not omit a major behavior it does contain. Write 120 to 220
words in plain prose with no TLA+ code.
\end{promptbox}

\paragraph{Intent description prompt (Claude version, verbatim).}
\begin{promptbox}[Intent Description]{intentcolor}
\ttfamily\small
You are writing a specification request, the way a person would describe a system they want
formally specified in TLA+. You are shown an existing TLA+ module, but your task is not to
describe that code. Recover the underlying intent and state it as a from-scratch request. State
what system or algorithm should be modeled, what it must do, and the key correctness properties
it must guarantee. Describe the goal and the requirements, not the implementation. Do not name
the module's variables, operators or PlusCal structure, and do not walk through the actions
step by step.
\end{promptbox}

\paragraph{Specification-generation prompt (verbatim).}
\begin{promptbox}[Specification Generation]{declcolor}
\ttfamily\small
You are a TLA+ specification engineer. You will receive a natural language description of a
system. Produce a complete, syntactically correct TLA+ specification that faithfully captures
the described behavior. The output must be a valid TLA+ module that passes the SANY parser.
Include the MODULE declaration, EXTENDS, CONSTANTS, VARIABLES, Init, Next and Spec. Include all
safety invariants and liveness properties described, and any fairness conditions mentioned.
Output only the TLA+ specification.
\par\medskip
System description: \{description\}
\end{promptbox}
\section{Datasheet for Datasets}
\label{sec:datasheet}

We document \Ours following the datasheet framework~\cite{datasheets}.

\textbf{Motivation.}
The dataset was created to measure natural-language to TLA\textsuperscript{+} specification
generation with an exact model-checker oracle, because prior resources grade by textual
similarity or by whether output parses, neither of which establishes correctness. It was
created by the authors.

\textbf{Composition.}
Each instance is a TLA\textsuperscript{+} specification with metadata. The dataset contains
1{,}300 specifications, 403 in a gold tier that both parses under SANY and model-checks under
TLC with a runnable configuration, and 897 in a silver tier that parses only. Each instance
carries a tier, a difficulty label, a category label for the gold tier, its source repository
with URL and path, the SHA-256 of its released file, and four model-written descriptions. Gold
instances additionally carry a runnable configuration, and instances that are trivial fixtures
carry a flag. The release also includes the model outputs graded in
Section~\ref{sec:experiments}. The specifications are drawn from 13 public repositories across
9 organizations. The dataset contains no personal data and, to our review, no offensive
content.

\textbf{Recommended splits.}
The gold and silver tiers are the primary split, and they differ in grading strength rather
than in sampling, since only the gold tier supports the semantic gate. The evaluation set used
in this paper is the system category of the gold tier with trivial fixtures removed, which
gives 100 specifications, and it is reconstructible from the released labels and flags. We
recommend no train and test split, since the dataset is an evaluation resource and should not
be trained on.

\textbf{Collection process.}
The specifications were crawled from public repositories. After removal of exact duplicates the
crawl yielded 1{,}614 candidate modules. An audited pipeline removed 314 modules that fail to
parse, are test scaffolding, fail at run time, are tooling rather than specifications, or are
near-duplicates. The remaining 1{,}300 form the release. The descriptions were generated by
GPT-5 and Claude Opus 4.5 with the prompts in Appendix~\ref{sec:prompts}.

\textbf{Preprocessing, cleaning, and labeling.}
The specification files are released unchanged from their sources, verifiable through the
recorded checksums. The cleaning removed test scaffolding and non-specification modules as
above. Trivial fixtures, which are valid specifications with a single reachable state, are
retained in the release, flagged in the manifest, and excluded from the evaluation set.
Difficulty labels were assigned to every specification and category labels to the gold tier.
Section~\ref{sec:difficulty} reports an inter-annotator reliability check on the difficulty
labels, and Section~\ref{sec:quality} reports a faithfulness audit of the descriptions. The
gold tier was defined by running SANY and TLC.

\textbf{Uses.}
The dataset is intended to evaluate models that generate TLA\textsuperscript{+} from
natural-language descriptions. It should not be used as training data, since training on an
evaluation benchmark defeats its purpose and contaminates future evaluations. The name-hidden
intent style additionally supports controlled studies of reliance on identifier names, and
Appendix~\ref{sec:rq-catalog} catalogs further uses the release enables. Users should note that
a reported correct rate depends on the grading regime, so a number from this benchmark should
be reported with the regime that produced it (Section~\ref{sec:envelope}).

\textbf{Distribution.}
The dataset is released on Zenodo under a persistent identifier and licensed under CC BY 4.0
for the descriptions and annotations. The specification files remain under the licenses of
their source repositories, which are recorded per instance.

\textbf{Maintenance.}
The authors maintain the dataset. Corrections and additional description sets will be released
as new versions under the same persistent identifier, so that a specific version can be cited.
Errors found in a label or a configuration will be corrected in a new version with the change
recorded in a changelog rather than by silently amending the archived one, so a result reported
against a given version stays reproducible.

\end{document}